\documentclass[11pt,cites,graphicx]{article}
\usepackage{epsfig}

\title{Fluctuations of water near extended hydrophobic and hydrophilic surfaces}

\author{Amish J. Patel and David Chandler \\[3mm]Department of Chemistry, University of California, Berkeley, California 94720 \\[1mm]}

\begin{document}

\maketitle

\begin{abstract}
We use molecular dynamics simulations of the SPC-E model of liquid water to derive probability distributions for water density fluctuations in probe volumes of different shapes and sizes, both in the bulk as well as near  hydrophobic and hydrophilic surfaces.  To obtain our results, we introduce a biased sampling of coarse-grained densities, which in turn biases the actual solvent density.  The technique is easily combined with molecular dynamics integration algorithms.  Our principal result is that the probability for density fluctuations of water near a hydrophobic surface, with or without surface-water attractions, is akin to density fluctuations at the water-vapor interface.  Specifically, the probability of density depletion near the surface is significantly larger than that in bulk.  In contrast, we find that the statistics of water density fluctuations near a model hydrophilic surface are similar to that in the bulk.
\end{abstract}

\section{Introduction}

According to theory, fluctuations of water density on sub-nanometer length scales obey Gaussian statistics~\cite{pratt_chandler, gaussian_ft, information_theory}, while those on larger length scales deviate significantly from this behavior~\cite{LCW}.  The deviations at large length-scales reflect the fact that water at standard conditions lies close to water-vapor coexistence \cite{LCW, fat_tail}.  A large enough repellent surface can therefore induce the formation of a water-vapor-like interface, and as such, the probability of water depletion is enhanced near such a surface.  In particular, the presence of this liquid-vapor-like interface facilitates large fluctuations in its vicinity compared to that in the bulk.  This perspective is at the heart of current ideas about hydrophobic effects \cite{DC_nature05, berne_rev09}. In this paper, we use molecular simulations to further examine the validity of this perspective.  We demonstrate marked similarities between water-vapor interfaces and water-oil interfaces for large enough oily surfaces with low radii of curvature.

The fact that a purely repulsive hydrophobic surface induces the formation of a vapor-liquid interface is well-established \cite{ LCW, stillinger, wallqvist, HC}. Dispersive attractions between the hydrophobic surface and water can mask this effect, as the attractions move the interface to a mean position immediately adjacent to the hydrophobic surface. This removes any significant presence of vapor on average \cite{HC, chou_dewet, chou05, ashbaugh01}. As a result, it is difficult to detect the presence of a water-vapor-like interface by considering only the mean behavior of water density profiles. Rather, the presence of this interface is more directly reflected in fluctuations away from those profiles \cite{DC_nature05, garde07, ball08cppc, mittal08, garde09prl, garde09, willard_fdy}, which motivates our focus here on the statistics of these density fluctuations.

Large length-scale water density fluctuations control pathways for assembling hydrophobically stabilized structures \cite{LCW, tWC, Willard_tWC, berne03, berne04, berne05_melittin, pgd06, shea08, pgd09}.  Several groups have studied the statistics of these fluctuations \cite{information_theory, garde96, garde00}.  This paper is distinguished from this earlier work by the fact that we report the statistics of large length-scale fluctuations, which are expected~\cite{LCW} to be fundamentally different from the small length-scale fluctuations studied in the earlier work.

The central quantity to be examined is the probability for finding $N$ water molecules in a sub-volume $\mathrm{v}$, $P_{\mathrm{v}}(N)$.  Hummer and Pratt and their co-workers introduced the idea of studying this function as a route to understanding solvation \cite{information_theory, garde96, garde06, pratt_book}. In pure solvent, large length-scale fluctuations are very rare.  These are the fluctuations of interest here.  Therefore, to obtain reasonable statistical information, we must employ some form of biased non-Boltzmann sampling~\cite{Frenkel_Smit, DCbook}. The simulation setup and system details are described in the next section, after which we explain the specific sampling method that we employ. Results are then presented, beginning with those for the length-scale dependence of fluctuations in bulk water, followed by a discussion of the effect of dispersive attractions between the water molecules and a hydrophobic surface. Finally, we contrast fluctuations near a hydrophobic surface with those near a hydrophilic surface.

Our results are consistent both with the Lum-Chandler-Weeks (LCW)~\cite{LCW} theory for the length-scale dependence of hydrophobic effects and with recent computer simulation studies \cite{garde09, berne08}.  In this sense, our results may seem unsurprising. This paper, however, is the first publication of computer simulation results demonstrating expectations of the LCW theory for fluctuations that deviate far from the mean in an atomistic model of water.  	

\section{Simulation Models}

Density fluctuations of SPC-E water \cite{spce} in the canonical ensemble at 298K can be quantified using the LAMMPS molecular dynamics (MD) simulation package.\cite{lammpsref}  
Specifically, we evaluate the probability, $P_{\mathrm{v}}(N)$, of finding $N$ water oxygens in a probe volume of interest, $\mathrm{v}$. A straight-forward canonical ensemble simulation, with an average water density, $\rho=1$ g/cm$^{3}$, would suppress large density fluctuations. To avoid this suppression, a particle-excluding field is introduced on one surface of the simulation box, with the box large enough that the net density is less than the bulk liquid density. Through this construction~\cite{bolhuis, tommy} we ensure that the bulk liquid remains in the center of the box, that it is at coexistence with its vapor phase, and that a  free liquid-vapor interface acts as a buffer in the event of a large density fluctuation.  Solutes are placed near the center of the box, deep within the bulk liquid and far from the free water-vapor interface.

To model a hydrophilic solute, we consider water molecules within a particular sub-volume of bulk water, specifically a sub-volume of dimensions $3\times24\times24$ {\AA}$^3$.  Taking a configuration of equilibrated bulk liquid, we immobilize those water molecules, which are within that considered sub-volume.  The immobilized molecules are then the solute.  Since the surface of this idealized hydrophilic solute is made up of fixed water molecules in a configuration that is typical of bulk water, the surface is unlikely to disrupt the hydrogen bond network of the neighboring solvent.

To model a hydrophobic solute, we replace the fixed waters in our model of a hydrophilic solute with methane-like oily particles.  These oily particles are uncharged and interact with the surrounding solvent water molecules via a standard water-methane Lennard-Jones potential.  The resulting potential energy field  expels water from the volume occupied by the solute.  It also attracts water with dispersive interactions. The solute constructed in this way and a typical configuration of solvent water are shown in Fig. 1. 

In order to study the effect of dispersive attractions on water-density fluctuations, the Lennard-Jones (LJ) pair potential between the hydrophobic solute and the solvent is split into repulsive and attractive parts using the Weeks-Chandler-Andersen (WCA) prescription. \cite{wca} The role of attractions can then be examined systematically with a scaling parameter $\lambda$,
\begin{equation}
u_{\lambda}(r)=u_{0}(r)+\lambda \Delta u(r), 
\end{equation}
where $u_0(r)$ and $\Delta u(r)$ are the WCA repulsive and attractive branches of the LJ potential, respectively, $\sigma=3.905${\AA} and $\epsilon=0.118$Kcal/mol were used as the LJ parameters for the oily particles \cite{jorgensen}, and Lorentz-Berthelot mixing rules were used to obtain water-solute interaction parameters.

\begin{figure}[htbp] 
\begin{center} 
\includegraphics[scale=0.4]{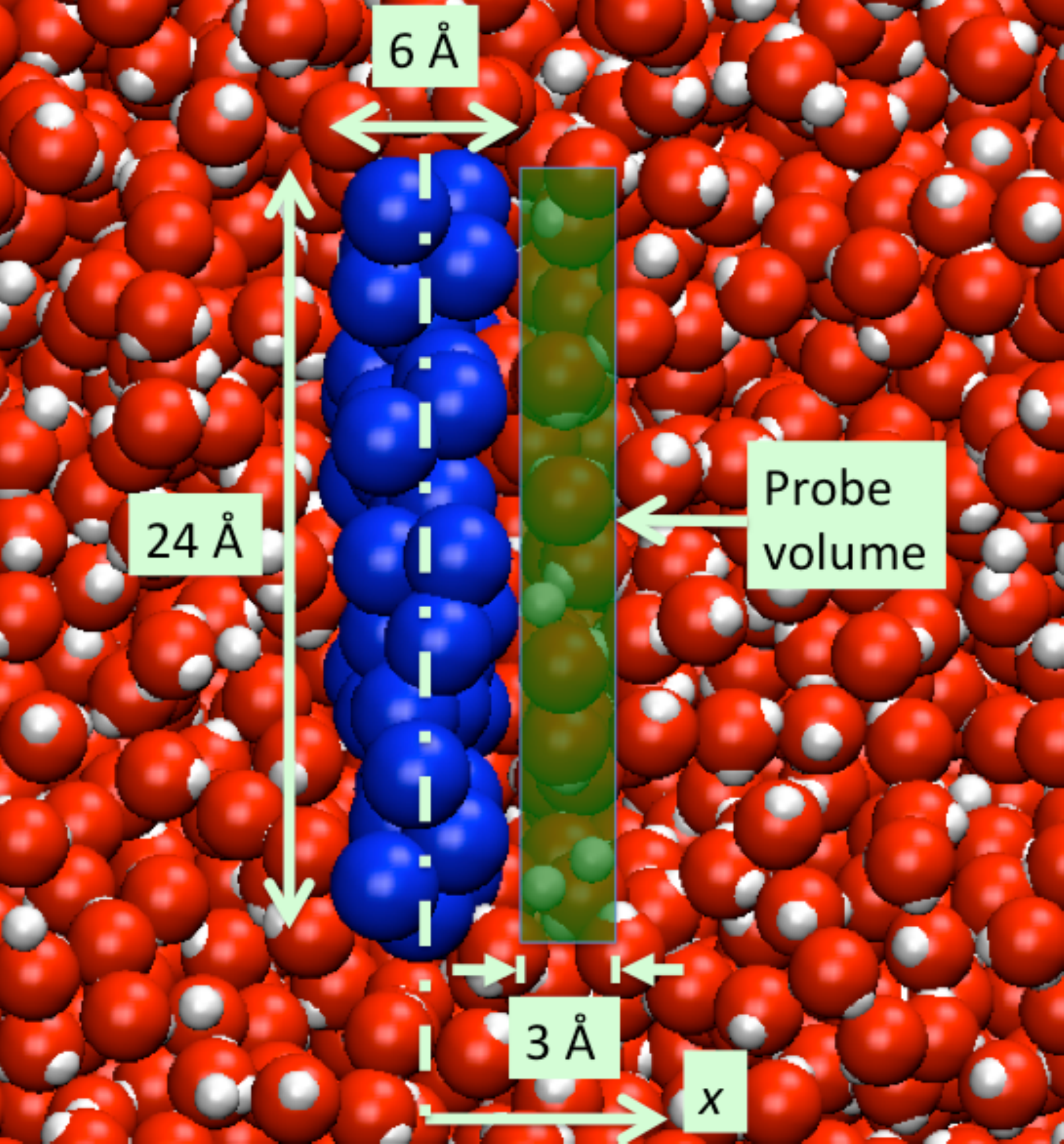} 
\caption{A typical probe volume, $\mathrm{v} = (3 \times 24 \times 24 )$ {\AA}$^3$, is shown here adjacent to a hydrophobic surface (blue particles).  The rendered water molecules (red and white) are in a typical equilibrium configuration taken from one of our simulations.}
\label{default}
\end{center}  
\end{figure} 

\section{Umbrella sampling}

Since we have chosen molecular dynamics to probe our system, it is convenient to use a biasing umbrella potential that produces continuous forces.  The potential we use is a function of the entire set of water oxygen co-ordinates: $\{\mathbf{r}_{i}\}$, $i=1,2,...M$. The number of molecules, $N$, in a specific volume, v, is not a continuous function of $\{\mathbf{r}_{i}\}$, but the biasing potential we use to influence this number is a continuous function of these variables. In particular, we focus on the coarse-grained particle number:

\begin{equation}
\hat{N}(\{\mathbf{r}_{i}\}, \mathrm{v})=\int d\mathbf{r} \sum_{i=1}^{M}  \Phi(\mathbf{r}-\mathbf{r}_{i}) h_{\mathrm{v}}(\mathbf{r}),
\end{equation}
where $h_{\mathrm{v}}(\mathbf{r})=1$ or 0, depending on whether or not $\mathbf{r}$ is in volume $\mathrm{v}$, $\mathbf{r}$ has Cartesian coordinates $x,y,z$, and
\begin{equation}
\Phi(\mathbf{r}) = \phi(x)\phi(y)\phi(z)
\end{equation}
with $\phi(x)$ being a normalized, truncated and shifted Gaussian-like distribution, 
\begin{equation}
\phi(x) \propto [\exp(-x^{2}/2\xi^{2})-\exp(-r_{\mathrm{c}}^{2}/2\xi^{2})] \theta(r_{\mathrm{c}}-|x|).
\end{equation}
The proportionality constant is the normalization constant, $\theta(x)$ is the heavy-side step function, and we have chosen the coarse-graining length $\xi$ to be 0.1{\AA} and the cut-off length $r_{\mathrm{c}}$ to be 0.2{\AA}.

In the limit $\xi \rightarrow 0$, the dynamical variable $\hat{N}(\{\mathbf{r}_{i}\}, \mathrm{v})$ is the actual number of water molecules in the volume v. But for finite $\xi$, $\hat{N}(\{\mathbf{r}_{i}\}, \mathrm{v})$ is a continuous and differentiable function of $\{\mathbf{r}_{i}\}$.  We construct the biasing potential with this variable.  In particular, we let
\begin{equation}
U(\{\mathbf{r}_{i}\};\kappa, \eta)=\frac{\kappa}{2} \left[ \hat{N}(\{\mathbf{r}_{i}\},\mathrm{v})-\eta \right]^{2},
\end{equation}
where $\kappa$ is a positive constant.  We have found it convenient to use $\kappa=0.25$ kcal/mol. Simulating our system in the presence of this umbrella potential allows us to bias the system towards configurations with $\hat{N}(\{\mathbf{r}_{i}\}, \mathrm{v})$values near $\eta$.  When $\eta$ differs substantially from the mean value for the dynamical variable, these configurations will be very improbable in the unperturbed system.  Nevertheless, these configurations can be accessed reversibly through a series of simulations that slowly change the control variable $\eta$. 

By influencing the coarse-grained number of particles in the probe volume, the control variable $\eta$ also influences the actual number of water molecules in that volume.  We have picked a small value of the coarse graining length $\xi$ to ensure that the latter influence is significant.  As a result, with a series of simulations with different values of $\eta$, histograms for both the coarse grained number and the actual number can be collected and then unbiased and stitched together within the framework of the weighted histogram analysis method (WHAM) ~\cite{wham1,wham2,roux_wham}.  This procedure yields the joint distribution function that the coarse grained particle number, $\hat{N}(\{\mathbf{r}_{i}\}, \mathrm{v})$, has value $\tilde{N}$ and the actual particle number is $N$. The joint distribution, $P_{\mathrm{v}}(N,{\tilde{N}})$, can then be integrated to give the distribution of interest, $P_{\mathrm{v}}(N)$.  With $P_{\mathrm{v}}(N)$  known, the free energy of solvation of a ``hard'' probe cavity of volume v, $\Delta\mu_{\mathrm{v}}$, is also known because \cite{information_theory}
\begin{equation}
\beta\Delta\mu_{\mathrm{v}}=-\ln P_{\mathrm{v}}(0)
\end{equation}
where $k_{\mathrm{B}}\beta=1/T$ is inverse temperature and $k_{\mathrm{B}}$ is Boltzmann's constant.

\begin{figure}[htbp] 
\begin{center} 
\includegraphics[scale=1.0]{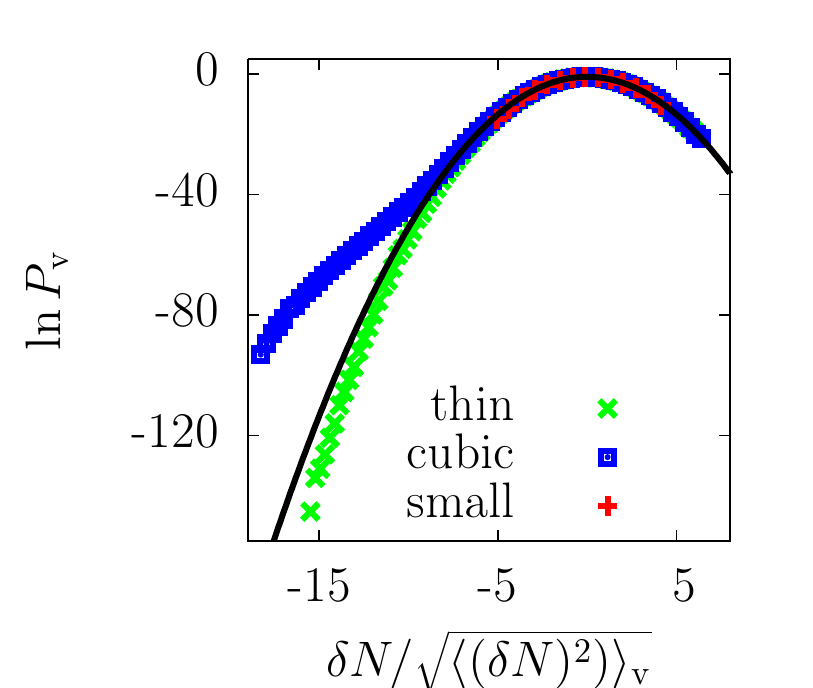} 
\includegraphics[scale=1.0]{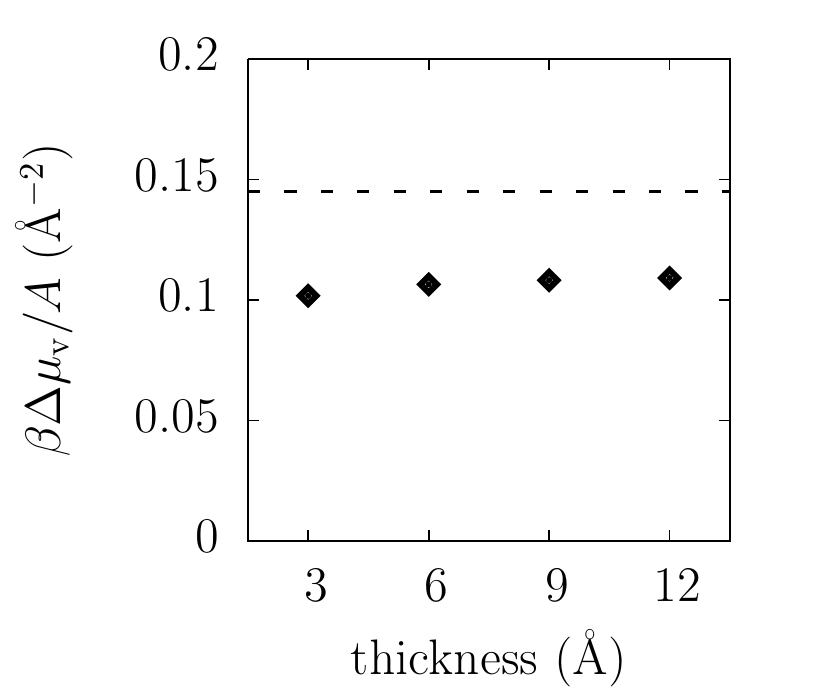} 
\caption{(a)  Probability distribution of finding $N$ water oxygens in a probe volume $\mathrm{v}$ in bulk water for a ``small'' cubic $\mathrm{v}= (6\times6\times6)$ {\AA}$^3$, a larger ``cubic'' $\mathrm{v}= (12\times12\times12)$ {\AA}$^3$ and a ``thin'' $\mathrm{v}= (3\times24\times24)$ {\AA}$^3$. The solid line refers to the Gaussian distribution with the same mean and variance; $\delta N=N-\langle N\rangle_{\mathrm{v}}$ refers to the instantaneous fluctuation in $N$ from its mean, $\langle N\rangle_{\mathrm{v}}$.  (b) The solvation free energy, $\Delta \mu_{\mathrm{v}}$, in units of $k_{\mathrm{B}} T$, per unit surface area, $A$, for probe cavities with different thicknesses and square cross-sections, but the same large volume $[(12$\AA$^3)=1728$\AA${^3} ]$. The dashed line indicates the value of the the surface tension of SPC-E water.}
\label{default}
\end{center}  
\end{figure} 

\section{Results}
\subsection{Length-scale dependence of density fluctuations in bulk water}  
In Fig. 2, we show results for $P_{\mathrm{v}}$($N$) in different probe volumes in bulk water.  Because $P_{\mathrm{v}}(N)$ is Gaussian for molecularly sized probe volumes~\cite{information_theory}, we present these results in comparison with Gaussians of the same mean, $\langle N \rangle _{\mathrm{v}}$, and the same variance, $\langle ( \delta N)^2 \rangle _{\mathrm{v}}$.  For small deviations from the mean, $P_{\mathrm{v}}(N)$ is essentially Gaussian, and for small volumes v, only small deviations from the mean $N$ are possible.  For large v, however, in the wings of the distribution for small $N$, the distribution differs markedly from Gaussian.  In pure water, the chance of observing these deviations is negligible, less than one part in many powers of ten.  On the other hand, these deviations become accessible and even dominant near a sufficiently large and repellent solute particle.  This is true because in these wings of the distribution the free energy to reduce $N$, namely $-k_{\mathrm{B}}T \ln \left[ P_{\mathrm{v}}(N) \right]$, can vary linearly or sub-linearly with $N$.  See Fig. 2a for the case of a large cubic probe volume.  As a result of this behavior, the introduction of a potential energy, which scales linearly with the number of particles it couples to, can favor low $N$ to the point where low values become the most probable values.  It is this shift in the distribution, which can occur only for large v, that is responsible for the large length-scale hydrophobic effects~\cite{LCW}.

In the case of a large but thin rectangular volume, the wings of $P_{\mathrm{v}}(N)$ also exhibit a deviation from Gaussian behavior.  However, in this case, the distribution lies below the Gaussian. 
Despite the fact that both the large cubic volume and the thin rectangular volume have the same volumetric size, the probabilities for emptying the cubic and the thin probe volumes differ by $\sim25$ orders of magnitude.  This behavior can be rationalized within the context of large length-scale solvation behavior being governed by the energetics of interface formation.  The thin volume has a much greater surface area than the cubic volume.  Indeed, the $\Delta\mu_{\mathrm{v}}$ computed from Eq. (6) for several probe volumes with fixed volumetric size but varying thickness confirms this rationalization, as shown in Fig. 2b.

\begin{figure}[htbp] 
\begin{center} 
\includegraphics[scale=1.0]{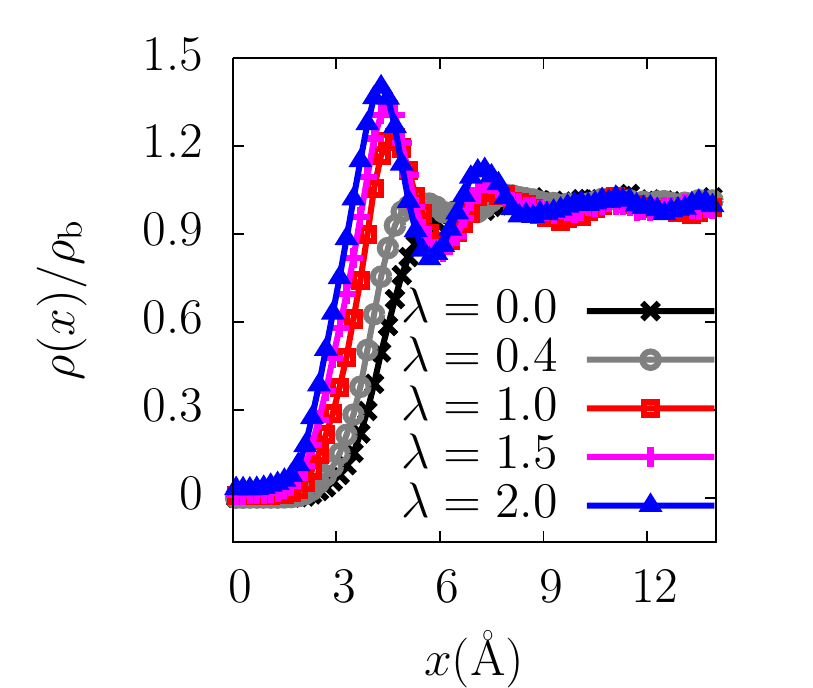} 
\caption{Mean water density $\rho(x)$, relative to its bulk liquid value, $\rho_{\mathrm{b}}$, perpendicular to extended hydrophobic solutes with different levels of dispersive attractions quantified by $\lambda$.}
\label{default}
\end{center}  
\end{figure}

\subsection{Water near hydrophobic surfaces and the effect of dispersive attractions}  
Figure 3 shows normalized mean densities as a function of the distance, $x$, from center of the idealized large flat hydrophobic solute (see Fig. 1).  Several strengths of dispersive attractions between the solute and water are considered.  See Eq. (1).  For $\lambda=0$, the mean density profile is sigmoidal, suggestive of a vapor-liquid interface. However, addition of a small amount of attraction results in a qualitatively different density profile. For $\lambda=0.4$, there is a maximum in the density profile accompanied by layering. Further increasing the attractions leads to a more pronounced maximum and layering.  This behavior is in accord with the predictions of LCW theory~\cite{HC}. Nevertheless, contrary to LCW theory, it has been suggested that a layered density profile implies an absence of a liquid-vapor-like interface near an extended hydrophobic surface with dispersive attractions to water.\cite{chou_dewet, chou_08}.  The source of this incorrect impression is that the mean density by itself is not an indicator of liquid-vapor-like interfaces.  Interfaces are relatively soft so that a weak perturbation can affect the location of the interface and thus the mean density profile while not destroying the interface. In other words, in order to fully appreciate the effect that a hydrophobic solute has on the surrounding solvent, one has to look at not only at the mean density but also at the density fluctuations.

\begin{figure}[htbp] 
\begin{center} 
\includegraphics[scale=1.0]{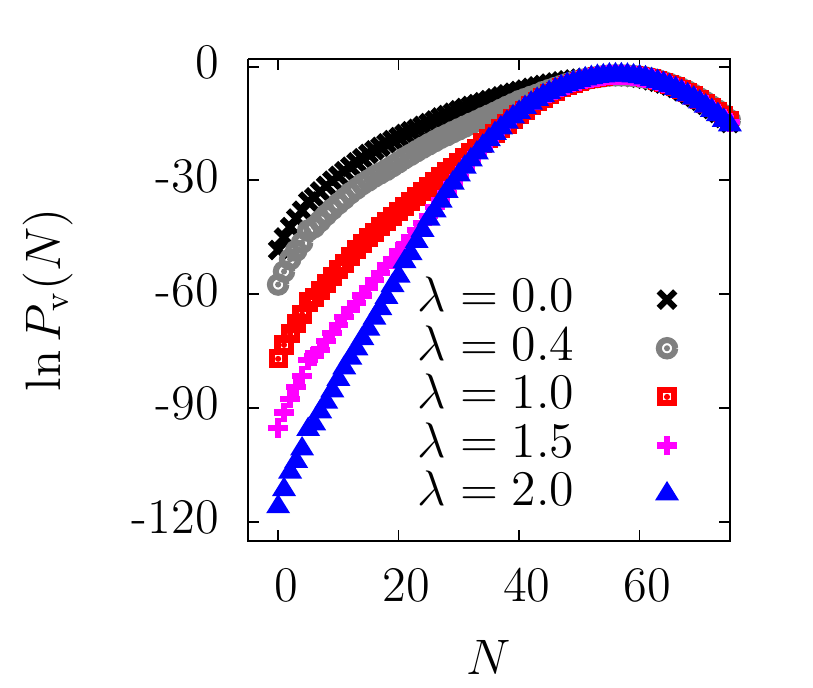} 
\includegraphics[scale=1.0]{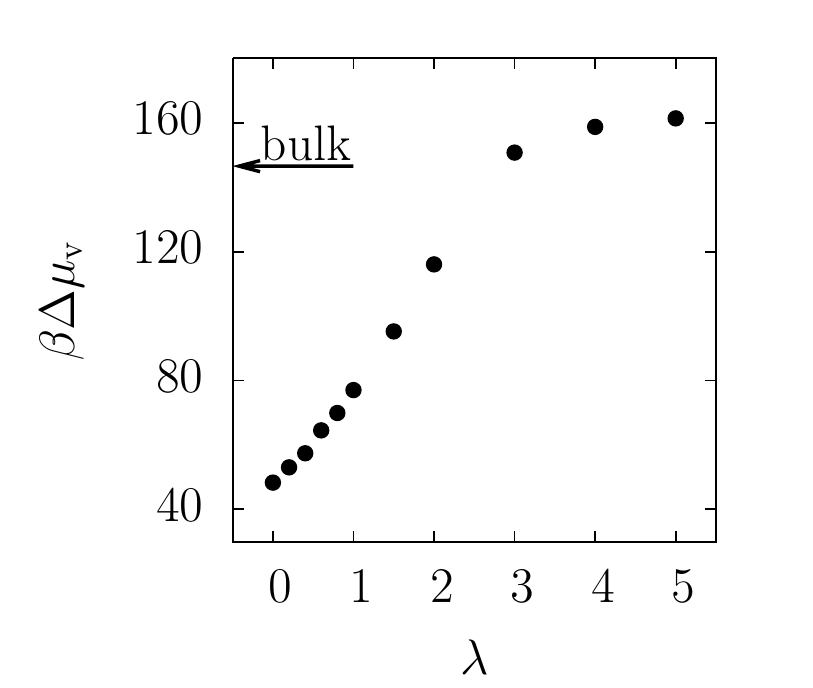} 
\caption{(a) $P_{\mathrm{v}}(N)$ for probe volumes near hydrophobic solutes with different $\lambda$ values. (b) The corresponding solvation free energies. For $\lambda<3$, it is easier to solvate $\mathrm{v}$ near the hydrophobic solute than it is to solvate $\mathrm{v}$ in bulk water.}
\label{default}
\end{center}  
\end{figure}

The statistics for these fluctuations can be obtained from the distribution of particle numbers in suitably chosen probe volumes.  Figure 4a shows $P_{\mathrm{v}}(N)$ distributions for the thin rectangular probe volume v=($3\times24\times24)$  {\AA}$^3$ placed between $x=5${\AA} and $x=8${\AA}. With this position, there is no overlap between the van der Waal's radii of solute particles and water molecules in $\mathrm{v}$. The distributions for $\lambda=0$ and $\lambda=0.4$ are similar, with the probability of density depletion slightly lower for the latter case, but still significantly higher than that in the bulk.

From Fig. 4b, we see that the free energy to empty this probe volume adjacent to the large hydrophobic solute with $\lambda=0.4$ is $57k_{\mathrm{B}}T$, whereas the free energy to empty this same v when it is in bulk and far from the hydrophobic surface is $147k_{\mathrm{B}}T$.  We can understand this large difference in terms of the free energetics of interface formation. As seen in Fig. 3, an extended hydrophobic solute with $\lambda=0$ leads to the formation of an interface due to the unbalancing of attractions in water. $P_{\mathrm{v}}$(0) for $\mathrm{v}$ adjacent to the solute is then essentially the probability to move the interface outwards by 3{\AA} from $x\simeq 5$ {\AA} to $x\simeq 8$ {\AA}. The free energetic penalty, $\Delta\mu_{\mathrm{v}}$ for this process is 48$k_{\mathrm{B}}T$ as shown in Fig. 4b. The corresponding $\Delta\mu_{\mathrm{v}}$ for $\lambda=0.4$ is 57$k_{\mathrm{B}}T$, only a 9$k_{\mathrm{B}}T$ increase on turning the attractions on to $\lambda=0.4$ and as much as 90$k_{\mathrm{B}}T$ less than that required to form interfaces. Hence, while the presence of a density maximum and layering at $\lambda=0.4$ might lead one to question the presence of an interface, the fluctuations in density or the ease with which a volume near the hydrophobic surface can be vacated leave no doubt as to its presence. Furthermore, Fig. 4b shows that one has to ramp up the attractions to nearly 3 times the value of typical dispersive attractions for the solvation free energy near a hydrophobic solute to equal that in the bulk.

\begin{figure}[htbp] 
\begin{center} 
\includegraphics[scale=1.0]{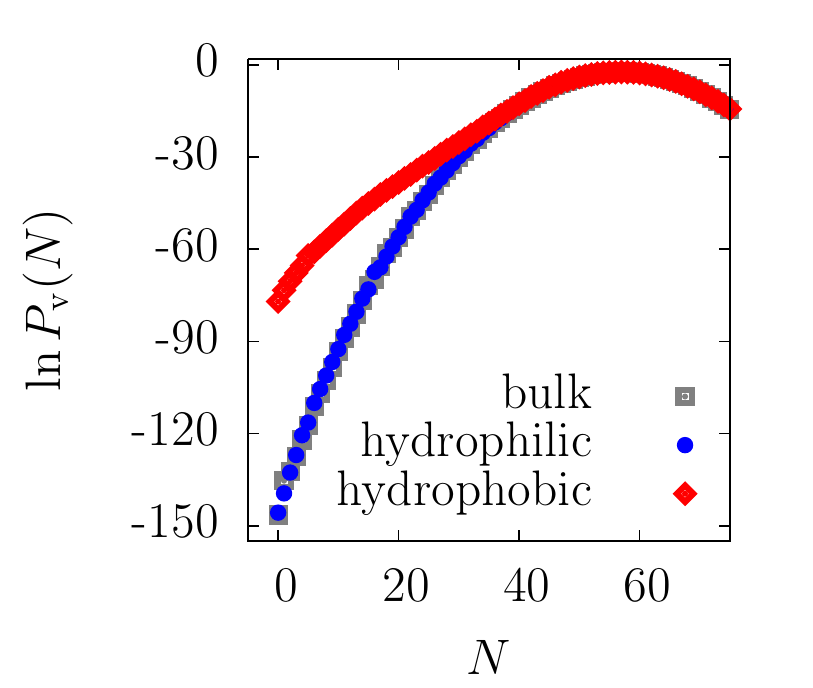} 
\includegraphics[scale=1.0]{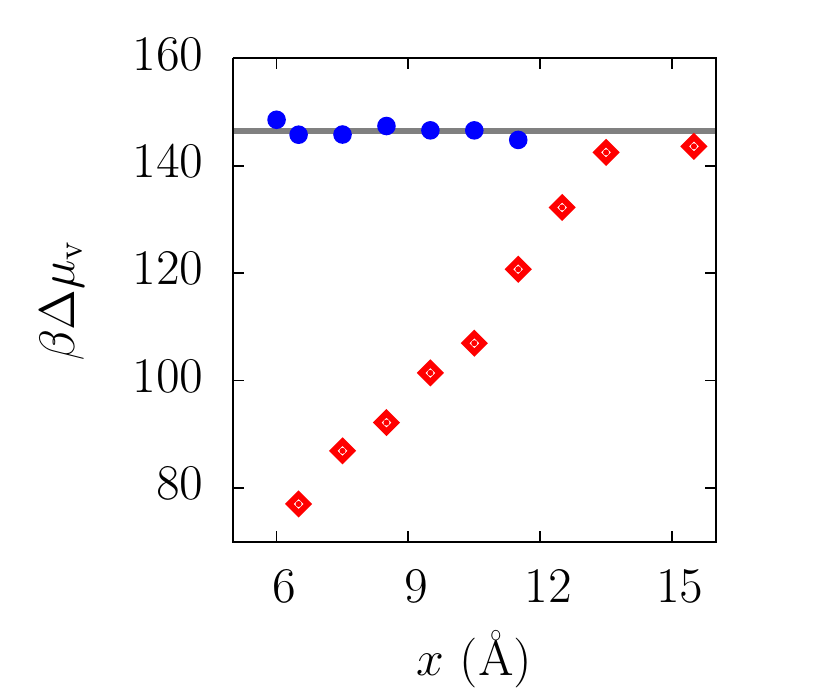} 
\caption{(a) $P_{\mathrm{v}}(N)$ for $\mathrm{v}=(3\times24\times24)$  {\AA}$^3$ near a hydrophobic solute, near a hydrophilic solute, and in bulk. (b) The change in solvation free energy of the probe volume $\mathrm{v}$ as it is moved away from the solute.}
\label{default}
\end{center}  
\end{figure}

\subsection{Fluctuations near hydrophobic and hydrophilic surfaces} $P_{\mathrm{v}}(N)$ for the probe volume $\mathrm{v}= (3\times24\times24)$  {\AA}$^3$, near the hydrophobic solute ($\lambda=1$) is compared with that for the probe volume near the hydrophilic solute in Fig. 5a.  $P_{\mathrm{v}}(N)$ near the hydrophilic solute (circles) is nearly identical to the $P_{\mathrm{v}}(N)$ for $\mathrm{v}$ in the bulk. (squares) On the other hand, the probability of density depletion for $\mathrm{v}$ near the hydrophobic solute (diamonds) is significantly higher. In Fig. 5b, we show the solvation free energy, $\Delta\mu_{\mathrm{v}}$, of the probe cavity as it is moved away from the solute, as a function of the distance between the center of the solute and the center of the cavity volume. While $\Delta\mu_{\mathrm{v}}$ for $\mathrm{v}$ near the hydrophilic solute (circles) remains equal to that for $\mathrm{v}$ in the bulk (solid line), $\Delta\mu_{\mathrm{v}}$ for $\mathrm{v}$ near the hydrophobic solute (diamonds) increases monotonically as $\mathrm{v}$ is moved away from the solute and eventually plateaus to the bulk $\Delta\mu_{\mathrm{v}}$. Our results indicate that the hydrophobic surface affects density fluctuations in the water at a distance of up to $\sim10${\AA}. These results are in agreement with recent simulation studies, reporting the free energy of solvating a molecularly sized WCA cavity near hydrophobic surfaces \cite{garde09} and also the potential of mean force for bringing two hydrophobic plates close together. \cite{berne08} 

These results bear directly on nano-scale assembly. In particular, we have shown that the probability of water density depletion near a hydrophobic surface is significantly enhanced.  When two such hydrophobic surfaces approach each other, at a particular separation, the liquid between them is sufficiently destabilized to make drying and hydrophobic assembly kinetically accessible.  In contrast, density fluctuations near a hydrophilic surface are identical to those in the bulk and the vapor phase is not stabilized by the presence of a hydrophilic surface.  Hydrophobic and hydrophilic surfaces thus differ fundamentally in the way they affect the fluctuations of water molecules in their proximity. It is not the mean density, but rather the statistics of fluctuations that is most important.

\section{Acknowledgement}

We thank Patrick Varilly for help with incorporating the methods discussed in this paper into the LAMMPS MD package and for pertinent suggestions. We would also like to acknowledge Adam Willard for helpful discussions as well as Shekhar Garde and Gerhard Hummer for sharing their work on similar issues prior to publication. This research was supported by NIH grant no. R01-FM078102.


\end{document}